\documentstyle[twocolumn,aps,prb,epsf]{revtex}
\begin{document}
\preprint{NUSc/96-03}
\title{
Comment on ``Predominantly Superconducting Origin ...", by N. Miyakawa, et al.}

\author{R.S. Markiewicz and C. Kusko$^{*}$} 

\address{Physics Department and Barnett Institute, 
Northeastern U.,
Boston MA 02115}
\maketitle


\narrowtext

\begin{figure}
\leavevmode
   \epsfxsize=0.33\textwidth\epsfbox{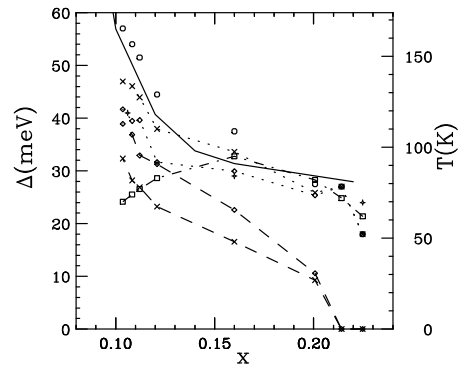}
\vskip0.5cm 
\caption{Tunneling gaps in BSCCO.
Circles = net tunneling gap, $\Delta$; dotted (dashed) lines = two estimates for
$\Delta_s$ ($\Delta_p$); squares = $T_c$; $+$'s = $10I_cR,$ where $I_cR$ is the
average $I_cR$ product\protect\cite{tu7}; solid line: from 
Ref.~\protect\cite{MKK}. }
\label{fig:1}
\end{figure}

Recent studies\cite{tu4,tu1,tu3} have found that the tunneling gap 
in Bi$_2$Sr$_2$CaCu$_2$O$_8$ (BSCCO) scales with the pseudogap, and not with
the superconducting $T_c$, as the doping is varied, Fig.~\ref{fig:1}.  
Miyakawa, et al.\cite{tu7} provide evidence that this gap is solely
associated with superconductivity and superconducting fluctuations, thereby 
contradicting scenarios in which the pseudogap is associated with competing 
magnetic or density wave instabilities\cite{KaSch,Zhang5,Laugh,WeL,Pstr,RAK}.
Here, we reanalyze the data of Ref.~\cite{tu7}, and show that while
fluctuation effects are important, there appears to be a second,
non-superconducting component of the gap which increases strongly with
underdoping.  We discuss three points raised by Miyakawa, et al. 

(1) {\it There is a single gap at all dopings.}  This is actually consistent 
with a number of two component theories.  In our model\cite{MKK}, the dominant 
gap, at the saddle point $(\pi ,0)$, is the vector sum of its components:
\begin{equation}
\Delta =\sqrt{\Delta_s^2+\Delta_p^2}, 
\label{eq:1}
\end{equation}
where $\Delta_s$ is the superconducting gap and $\Delta_p$ a competing 
pseudogap.  This form was also proposed phenomenologically by Loram, et 
al.\cite{Lor}.  In contrast, near $(\pi /2,\pi /2)$ our model predicts a gap 
near the Fermi level due to superconductivity only.  For a d-wave gap, there is 
no peak near $(\pi /2,\pi /2)$, and this feature would be difficult to detect in
tunneling.  However, Panagopoulos and Xiang\cite{PX}
have determined that {\it the slope of the gap near the gap zero at $(\pi /2,\pi
/2)$ scales with $T_c$, and not with the gap near $(\pi ,0)$!}

(2) {\it There is a dip feature above the tunneling peak, which scales with the 
gap}.  Since the dip is due to reduced scattering in the superconducting 
state\cite{CoCo}, it provides a measure of $\Delta_s$.  Thus, Fig. 2 of 
Ref.~\cite{tu7} shows that $\Delta_s/\Delta$ is not constant, since the dip 
minimum systematically evolves with doping.   In Fig.~\ref{fig:1} we provide 
two estimates of $\Delta_s$ and (from Eq.~\ref{eq:1}) $\Delta_p$: the $\times$'s
(diamonds) are found by assuming that the dip minimum (upper edge of the dip) is
exactly proportional to $\Delta_s$ and that $\Delta_p$ vanishes in the most 
overdoped sample.  The upper edge probably provides a better estimate, since the
lower edge of the dip can be obscured by the upper edge of the tunneling peak.  

(3) {\it The Josephson $I_cR$ product is found to increase in the underdoped 
regime} ($+$'s in Fig.~\ref{fig:1}), scaling well with the upper-edge estimate 
of $\Delta_s$.  Thus, there are clearly enhanced two-dimensional fluctuations
in the underdoped regime, where the ratio $\Delta_s/T_c$ increases by about a 
factor of two (Fig.~\ref{fig:1}).  However, at the lowest 
doping, the non-superconducting component of the gap has grown to equal
$\Delta_s$ in magnitude.  The solid curve in Fig.~\ref{fig:1} represents a 
universal pseudogap curve for YBa$_2$Cu$_3$O$_{7-\delta}$, La$_{2-x}$Sr$_x$CuO$
_4$ (LSCO)\cite{BatT} and BSCCO\cite{MKK}.  It extrapolates to a very large 
pseudogap, $\sim 300$meV ($\sim 2.2J$, where $J$ is the exchange energy) at 
half filling. This large zero-doping-limit gap is likely to be predominantly due
to $\Delta_p$.

Publication 744 of the Barnett Institute.

{\bf $*:$} On leave of absence from Inst. of Atomic Physics, Bucharest, 
Romania


\begin{references}
\bibitem{tu4}M. Oda, 
et al., 
Physica C{\bf 281}, 135 (1997).
\bibitem{tu1}Ch. Renner, 
et al.,
Phys. Rev. Lett. {\bf 80}, 149 (1998).
\bibitem{tu3}N. Miyakawa, 
et al.,
Phys. Rev. Lett. {\bf 80}, 157 (1998).
\bibitem{tu7}N. Miyakawa, 
et al.,
unpublished (cond-mat/9809398).
\bibitem{KaSch}A. Kampf and J. Schrieffer, Phys. Rev. B{\bf 41}, 6399 (1990).
\bibitem{Zhang5}S.-C. Zhang, Science {\bf 275}, 1089 (1997).
\bibitem{Laugh}R.B. Laughlin, J. Phys. Chem. Sol. {\bf 56}, 1627 (1995).
\bibitem{WeL}X.-G. Wen and P.A. Lee, Phys. Rev. Lett. {\bf 76}, 503 (1996).
\bibitem{Pstr}R.S. Markiewicz, Phys. Rev. B{\bf 56}, 9091 (1997).
\bibitem{RAK}R.A. Klemm, unpublished.
\bibitem{MKK}R.S. Markiewicz, C. Kusko and V. Kidambi, unpublished 
(cond-mat/9807068).
\bibitem{Lor}J.W. Loram, 
et al.,
J. Supercond. {\bf 7}, 243 (1994).
\bibitem{PX}C. Panagopoulos and T. Xiang, Phys. Rev. Lett. {\bf 81}, 2336 
(1998).
\bibitem{CoCo}D. Coffey and L. Coffey, Phys. Rev. Lett. {\bf 70}, 1529 (1993).
\bibitem{BatT}B. Batlogg, 
et al.,
Physica C{\bf 235-240}, 130 (1994).
\end{references}
\end{document}